# Supercontinuum generation in varying-dispersion and birefringent silicon waveguide

*Neetesh Singh[1*], Diedrik Vermulen[1], Alfonso Ruocco[1], Nanxi Li[1,2], Erich Ippen[1], Franz X Kärtner[1,3] and Michael R Watts[1]*

[1]*Research Laboratory of Electronics, Massachusetts Institute of Technology, Cambridge, MA 02139, USA;*
[2]*John A. Paulson School of Engineering and Applied Science, Harvard University, Cambridge, MA 02138, USA*
[3]*Centre for Free Electron Laser Science (CFEL)-DESY and University of Hamburg, Hamburg 22607, Germany*
*[*]neeteshs@mit.edu*

**Abstract:**
Ability to selectively enhance the amplitude and maintain high coherence of the supercontinuum signal with long pulses is gaining significance. In this work an extra degree of freedom afforded by varying the dispersion profile of a waveguide is utilized to selectively enhance supercontinuum. As much as 16 dB signal enhancement in the telecom window and 100 nm of wavelength extension is achieved with a cascaded waveguide, compared to a fixed dispersion waveguide. Waveguide tapering, in particular with increasing width, is determined to have a flatter and more coherent supercontinuum than a fixed dispersion waveguide when longer input pulses are used. Furthermore, due to the strong birefringence of an asymmetric silicon waveguide the supercontinuum signal is broadened by pumping simultaneously with both quasi-transverse electric (TE) and quasi-transverse magnetic (TM) mode in the anomalous dispersion regime. Thus, by controlling the dispersion for the two modes selective signal generation is obtained. Such waveguides offer several advantages over optical fiber as the variation in dispersion can be controlled with greater flexibility in an integrated platform. This work paves the way forward for various applications in fields ranging from medicine to telecom where specific wavelength windows need to be targeted.

## 1. Introduction

Supercontinuum (SC) generation is an important phenomenon due to its applications in optical frequency metrology and synthesis [1], ultra-short pulse generation [2], microwave photonics [3], hyperspectral light detection and ranging (LiDAR) [4], telecommunication [5] among others. Integrated supercontinuum sources are poised to become critical photonics elements as they are key components in integrated digital optical frequency synthesizers [6, 7], chemical sensors [8] and pulse compressors with compression down to the sub-10 fs level [9]. Recent developments in supercontinuum sources on various novel platforms provide clear indication of these interests. Some examples are, supercontinuum generation in integrated silicon nitride [10-13], aluminum nitride [14], chalcogenide [15, 16], indium gallium phosphide [17], silicon on germanium [18, 19], amorphous silicon [20], silica [21], high-index-doped silica [22], and lithium niobate waveguides [23]. In particular, silicon-on-insulator waveguides, when pumped near the low two-photon absorption window [24-29] in order to avoid nonlinear loss [30-32], are promising as octave spanning supercontinuum sources that can leverage the complementary metal oxide semiconductor (CMOS) industry.

Traditionally, supercontinuum generation in integrated devices is achieved by optimizing the cross-section of a waveguide to obtain a desired dispersion profile that is fixed for the entire propagation length. Despite demonstrations in the optical fibers [33-38], only recently numerical studies have been reported on supercontinuum enhancement in varying dispersion waveguides, in which the input pulse experiences varying dispersion along the propagation length [39-41].

In this work, some of the possible ways a waveguide can offer varying dispersion to enhance the supercontinuum signal and improve its coherence are explored. Especially, the cascaded, tapered, and highly birefringent silicon-on-insulator waveguides are studied. Significant amplitude enhancement (16 dB) with the cascaded dispersion waveguide at telecom C-band and 100 nm extension of the SC over its fixed dispersion counterpart are demonstrated. Furthermore, numerically a cascaded waveguide design that can improve significantly the *f*, *2f* and telecom signal of an octave spanning SC is shown, for use in the mode-locked laser self-referencing [6, 13]. The tapered waveguide is experimentally shown to have a flatter SC spectrum with the ability to maintain high coherence over a large bandwidth with longer pulses. The coherence is found to be maintained due to the continually changing dispersion of the taper thus avoiding the modulation instability based signal build-up before soliton fission occurs, which is inevitable in a fixed dispersion waveguide. Lastly, the birefringence of a waveguide to generate a broad supercontinuum by simultaneously pumping the waveguide with TE and TM modes is utilized. The flexibility in modifying the geometry of a waveguide for a desired dispersion (varying/fixed) gives integrated photonics a significant edge over optical fibers in the ability to tailor the SC. Thus, this work can potentially help improve the integrated supercontinuum devices where specific wavelength windows need to be targeted, for example, in bio-sensing and optical frequency metrology [29, 42, 43]. This work has also applications in long pulse coherent supercontinuum generation, and efficient self-referencing [43-45].

## 2. Cascaded waveguide

In this section a cascaded waveguide is discussed to demonstrate selective enhancement of the supercontinuum in comparison with a fixed width waveguide. The fixed width waveguide was designed to have a width of 920 nm, total height of 380 nm, slab thickness of 65 nm, and length of 1 cm. The cascaded waveguide consists of two sections, 920 nm and 1070 nm wide rib waveguides, each 5 mm long as shown in Figure 1(a). The cross-section of the waveguide is shown in Fig. 1(b, inset).

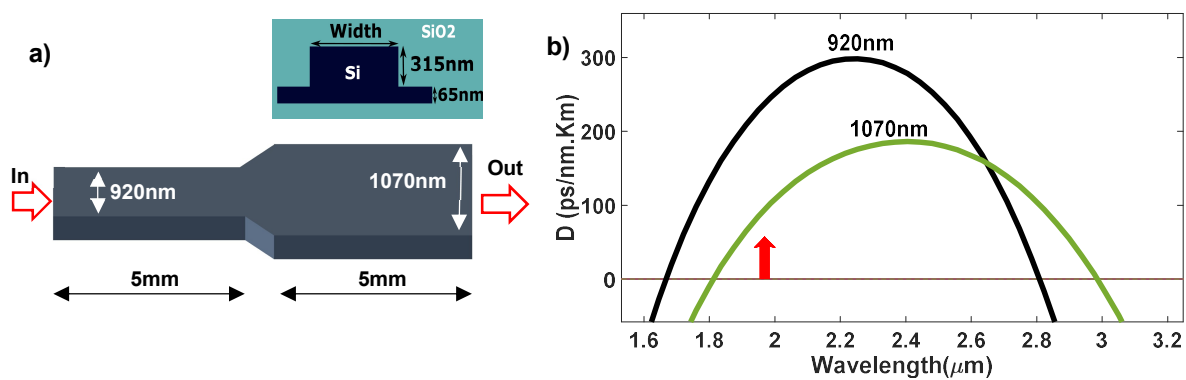

**Figure 1. (a)** An illustration of the cascaded waveguide with its cross-section (width is 920 nm/1070 nm). **(b)** The simulated dispersion curves of the cascaded waveguide are shown, with the red arrow indicating the pump wavelength at 1.95 μm.

The waveguides were fabricated on a 65 nm CMOS 300 mm platform at CNSE, SUNY, New York. The propagation loss of the cascaded waveguide and the other structures discussed in this paper is estimated to be between 1-1.5 dB/cm at 1.95 μm. The two sections of the cascaded waveguide are connected with a short adiabatic taper of 50 μm length. The dispersion curves are shown in Figure 1(b). The 1$^{st}$ zero dispersion wavelengths (ZDW) of the two sections of the cascaded waveguide are 1.65 μm and 1.8 μm. In both cascaded and fixed width waveguides inverse taper couplers of 200 μm length are used at the input and the output.

For characterization, the waveguides were pumped with an optical parametric oscillator (OPO) with pulse width of 250 fs and repetition rate of 80 MHz. The coupled peak power was 100 W, corresponding to a soliton number $N = 25$ for a group velocity dispersion (GVD) of -0.46 ps$^2$/m in the 920 nm waveguide at the pump wavelength, with the soliton fission length of 1.5 mm. The SC spectra for the fixed width and the cascaded waveguides are shown in Figure 2(a). The signals, especially around the telecom C-band and at 2.2 μm, are enhanced in the cascaded waveguide. The integrated power in the C-band is enhanced by 16 dB and the long wavelength signal is extended by 100 nm.

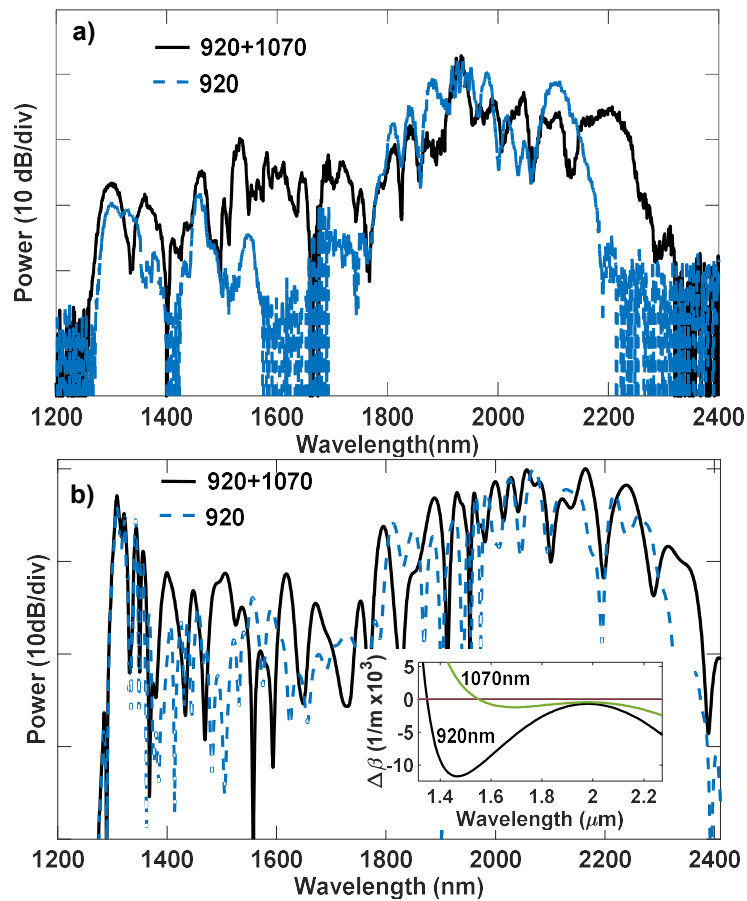

**Figure 2.** (a) The experimental, and (b) the simulated supercontinuum spectra for the fixed width (920 nm wide) waveguide (blue-dashed) and the cascaded waveguide (black-solid). The phase matching curves for the dispersive wave for the 920 nm and the 1070 nm wide section of the waveguide are shown in the inset.

The numerical solution of the nonlinear Schrodinger equation (NLSE) [46, 47] shown in Figure 2(b) matches to the experiment, with a slight shift in the enhancement region that can be attributed to the dimensional uncertainty. The dispersive wave phase matching (DWPM) curves (inset in Figure 2(b)) show the wavelength of dispersive wave generation (where the Δβ=0) for the two sections of the cascaded waveguide. The dispersive waves are generated in the 1st and the 2nd section at 1.34 μm and around 1.55 μm, respectively. In the 1st section the SC takes similar form as the SC in the fixed width waveguide (920 nm); however, as soon as the pulse enters the 2nd section (1070 nm), the fundamental solitons from the 1st section start ejecting energy into the dispersive waves, which are around 1550 nm [48, 49], thus increasing the signal strength of the C-band, while the solitons themselves simultaneously keep shifting to the longer wavelengths. The fundamental solitons shift to longer wavelengths (spectral recoil [49]), because the GVD in the 2nd section of the waveguide is lower than that in the 1st section, which increases their soliton number, determined by $N^2 = \gamma P T^2/\beta_2 = 1$ (for fundamental soliton) [46], where $\gamma$, $P$, $T$ and $\beta_2$ are the nonlinearity factor, power, temporal width of the pulse and the GVD, respectively. Thus, to become fundamental solitons again these higher order solitons shift to longer wavelength by 100 nm where the GVD is high, and also shed their energy to the dispersive waves.

These results indicate that the power in a spectral window of interest can be enhanced by cascading waveguides of different widths. Therefore, in the next section we show numerically a more compact device to enhance specific windows of an octave spanning supercontinuum which can be used, for example, in optical frequency synthesizer [6, 7].

*2.1 Three-section cascaded waveguide*
In the following, a 3 mm long device is shown that can be used to enhance the *f*, *2f* and 1550 nm region of the supercontinuum spectrum.

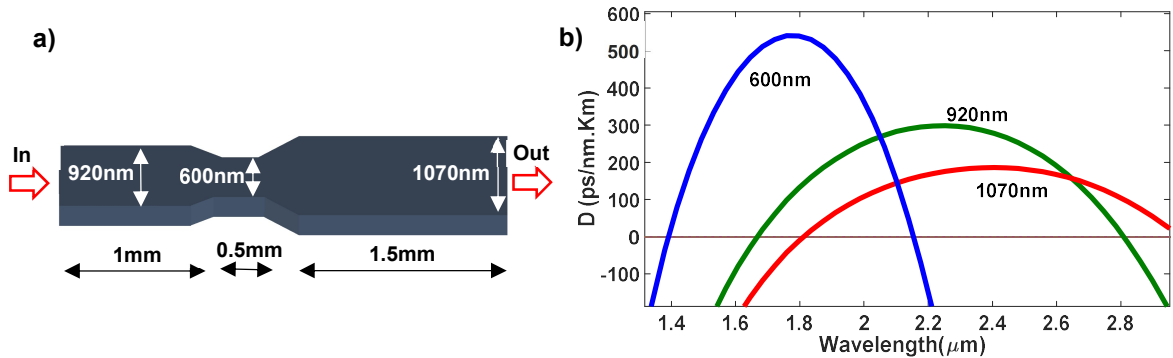

**Figure 3. (a)** The illustration of the three-section cascaded waveguide. **(b)** The dispersion curves for the cascaded waveguide: 1st section (600 nm wide – blue), 2nd section (920 nm wide – green), and 3rd section (1070 nm wide - red).

The three-section cascaded (3SC) waveguide is shown in Figure 3(a). The 1st section is 920 nm wide and 1 mm long, the 2nd section is 600 nm wide and 0.5 mm long, and the 3rd section is 1070 nm wide and 1.5 mm long, with total height and slab thickness as previously. The dispersion curve of each section is shown in Figure 3(b). The 1st ZDW of the 1st and the 3rd section are the same as in Figure 1(b), whereas for the 2nd section it is around 1.4 μm. In the numerical study a pulse of 100 fs width and a peak power of 100 W was coupled into the waveguide. The simulation results are shown in Figure 4. The SC spectrum is significantly stronger with the 3SC waveguide than for the fixed width waveguide of same length. The fixed width waveguide has ~

20 dB lower signal around 2.7 µm and about 7 dB weaker signal in the telecom widow compared to the 3SC waveguide. It must be noted that the fixed width of 920 nm was used for the above comparison because it generates broader SC spectra in the relevant wavelength region than the 1070 nm and 600 nm widths.

In the following, the propagation of the pulse through the 3SC waveguide is numerically studied. The light is launched in the 1st section in which the soliton number is 18, and the soliton fission length is 0.6 mm. After dispersive wave generation (~1.35 µm) around the soliton fission point the pulse propagates further in the 1st section for about 400 µm for the dispersive wave to build up enough energy before the whole pulse transitions into the 2nd section. In the 2nd section the fundamental soliton, arriving from the 1st section, starts phase matching to the dispersive wave (around 2.7 µm) as per the DWPM curves shown in the inset of Figure. 4. We observe that the solitons that phase match to the dispersive waves in the 1st and the 2nd section are at different wavelengths viz. at 1.8 µm for the 2nd section and 2.1 µm for the 1st section. The length of the 2nd section is optimized to increase the strength of the dispersive wave (~2.7 µm) relative to the strength of the dispersive wave in the 1st section. In the 3rd section the solitons generate dispersive wave around the telecom window while also shifting to longer wavelengths to become fundamental solitons again, as discussed above. We note that the lengths of different sections of the 3SC waveguide can be modified as per one's requirements, to vary the strength of the desired dispersive wave while keeping the total length device length fixed.

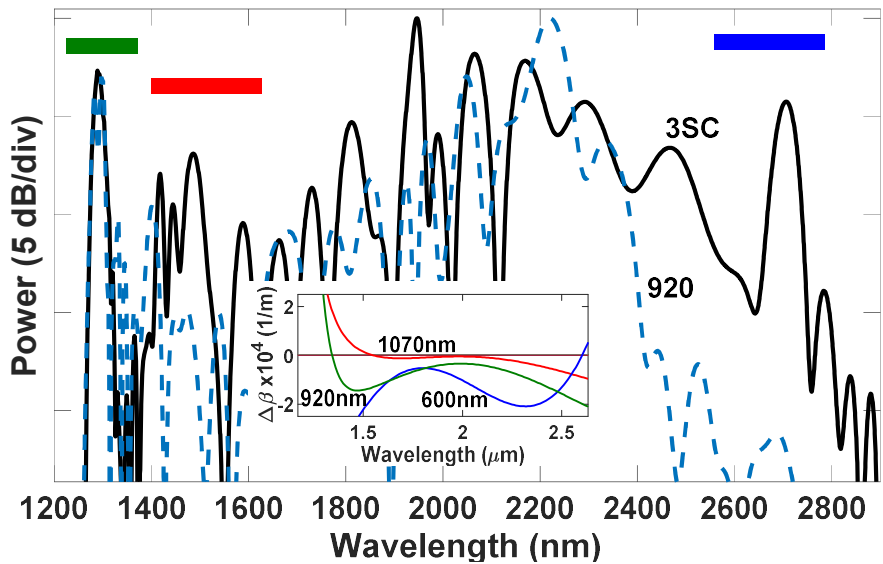

**Figure 4.** The simulated supercontinuum spectra of the three-section cascaded waveguide (black solid) and the fixed width (920 nm wide) waveguide (blue dash). The 1st, 2nd, and the 3rd section of the waveguide mainly generate the signal under the green, blue and red bar, respectively. The dispersive wave phase matching curves are shown in the inset.

## 3. Tapers

Tapers are another type of longitudinally dispersion varying waveguide that have been studied extensively in optical fibers [35-37, 41, 50]. The linearly increasing width waveguide is shown in Figure 5a (inset). The waveguide is 5 mm long and varies adiabatically from 500 nm to 1100 nm in width. The GVD variation along the length of the taper at the pump wavelength at 1.95 µm is shown in Figure 5a. The input signal is coupled near the 2nd ZDW of the waveguide and reaches at the output close to its 1st ZDW. The dispersion curves for the widths increasing in steps of 100 nm from the input to the output of the taper are shown in Figure 5(b). In the experiments the input pulse was launched using an OPO with the pulse width of 250 fs and a coupled peak power of 200 W. The results are shown in Figure 6(a) where the supercontinuum from the taper is compared to that of various fixed width waveguides (500 nm, 700 nm, and 1100 nm).

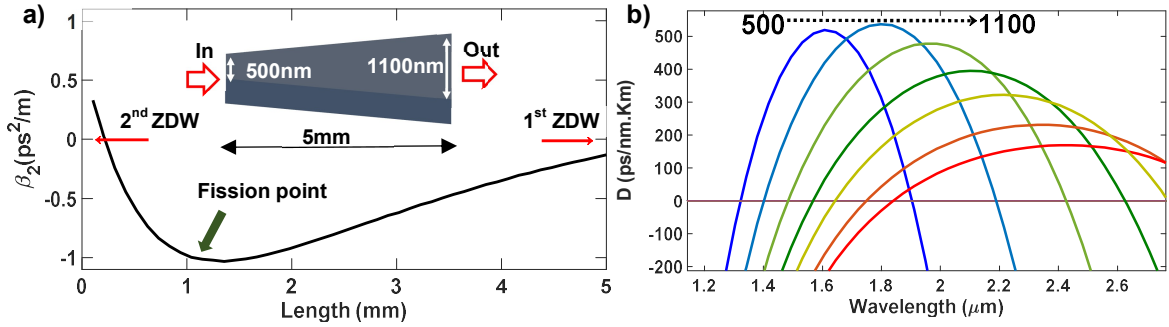

**Figure 5. (a)** Group velocity dispersion at the pump wavelength along the length of the increasing taper, along with the device illustration (inset). The soliton fission happens around 1.1 mm. **(b)** Dispersion curves from the start (blue – 500 nm) to the end (red – 1100 nm) of the taper.

The SC signal of the 500 nm wide waveguide is mainly based on self-phase modulation of the pulse since the pump is in the normal dispersion region. In the 700 nm wide waveguide the dispersive wave is generated at 1180 nm, and around the pump wavelength the soliton fission broadens the spectrum. Similarly, the spectral broadening in the 1100 nm wide waveguide is based on the soliton fission process. The supercontinuum generation in the taper is also based on soliton fission, and extends from <1200 nm to >2400 nm and is relatively flat in the long and the short wavelength regions of the SC. The amplitude of the spectrum varies within 5-7 dB from 1200-1850 nm and 3-5 dB from 2100-2400 nm. The simulations are shown in the Figure 6(b).

The reason for the flat spectrum with the taper is the shifting of the dispersive wave phase matching wavelength along the length of the taper due to the continuous change in width along the length of the taper [36, 37]. Beyond the soliton fission point (shown in Figure 5(a), at the width of 680 nm) the solitons keep on generating dispersive wave starting from 1.2 μm to longer wavelengths, while the solitons themselves keep shifting to longer wavelengths, thus helping to flatten the spectrum. We note that the pulse is launched into the normal dispersion region of the taper, this is to ensure that the soliton fission point (which is dependent on the pump power) is located near $d\beta_2/dL \approx 0$ (at 1.1 mm, see Figure 5(a)), causing the solitons to phase match to the dispersive waves in the short wavelength side of the pump rather than the long side. This is because the long side dispersive waves would be generated in the high propagation loss region of the SOI, as well as cannot be captured by our optical spectrum analyzers (OSA), Yokogawa AQ6375B.

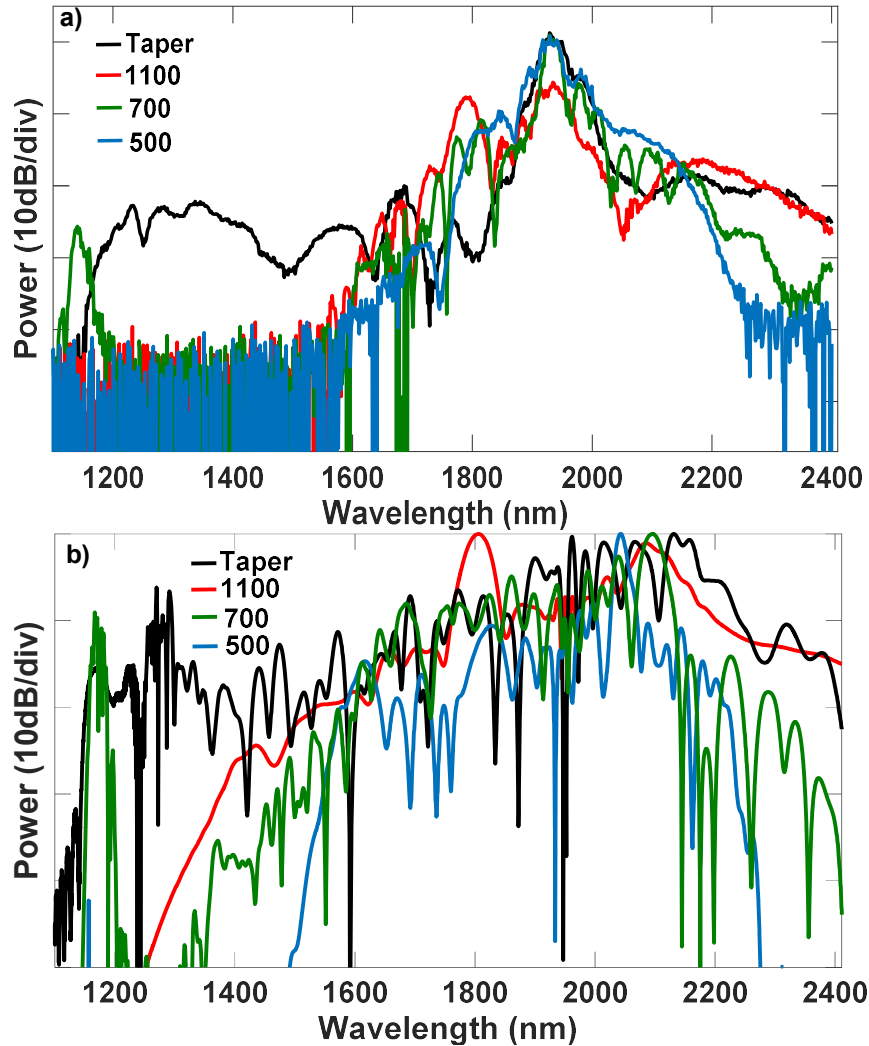

**Figure 6. (a)** Experimental and, **(b)** simulated supercontinua from increasing taper (black) and fixed width 500 nm (blue), 700 nm (green) and 1100 nm (red) wide waveguide.

*3.1 Increasing vs decreasing width taper*

In the following, the supercontinua from the increasing and decreasing width taper are compared. The pump parameters for the decreasing width taper are the same as for the increasing width taper discussed above. The experimental and numerical results are shown in Figure 7(a) and (b). We observe that the supercontinuum from the decreasing taper is not as flat as that of the increasing taper. This is because the soliton fission point in the decreasing taper is around 2.5 mm from the input where the width is 800 nm. This in turn causes the dispersive waves to be generated from 1.35 µm continuously to shorter wavelengths as indicated by the arrow (red) in Figure 7(c). As soon as the pulse reaches close to the point where the width of the taper is ≈ 650 nm the higher order dispersion terms of the waveguide suppress the short side dispersive waves, and the pulse starts phase matching to the dispersive waves at the long side of the pump, in the region not accessible by our OSA. Moreover, the solitons are travelling into the higher GVD region (see Figure 7(d)) along the decreasing taper thus they cannot shift, as in increasing taper, to longer wavelengths to remain fundamental solitons.

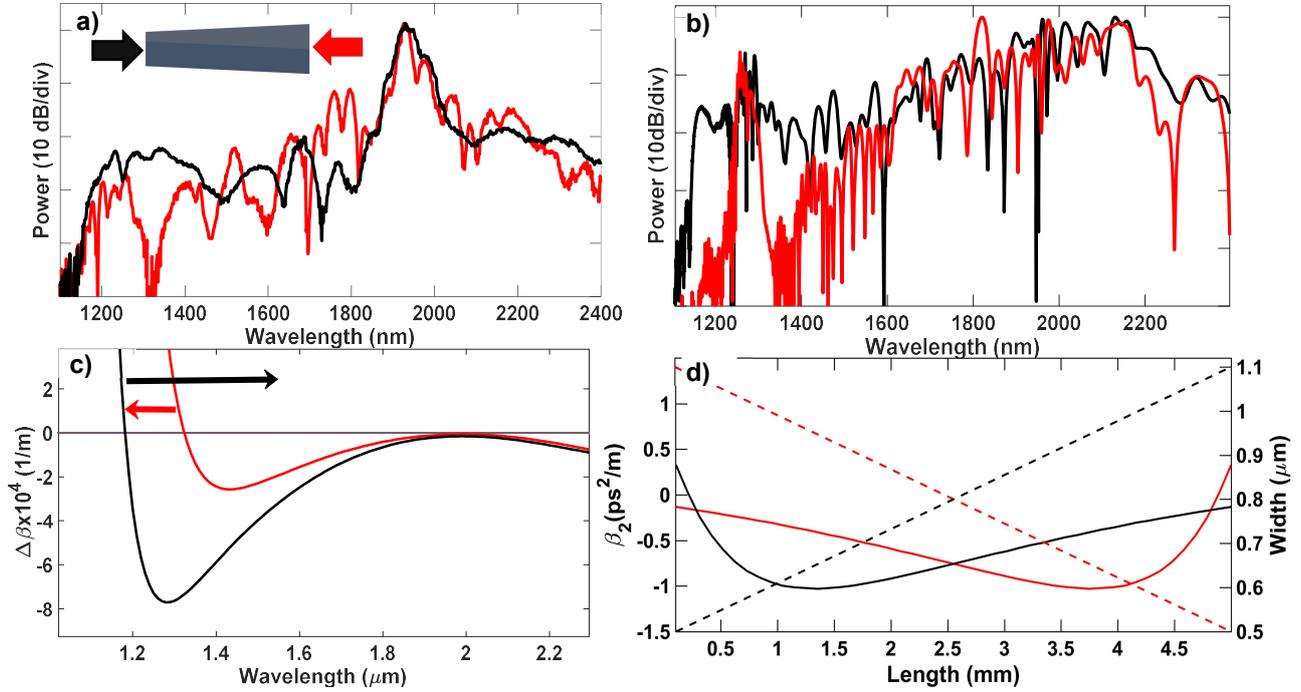

**Figure 7. (a)** Experimental and, **(b)** simulated supercontinua for the increasing width taper (black) and decreasing width taper (red). **(c)** Dispersive wave phase matching curves for the decreasing (red) and increasing taper (black), with arrows indicating shifting dispersive wave after soliton fission point. **(d)** The variation of the GVD and the width along the length of the taper for the increasing (black) and decreasing (red) taper. The dispersion and width vertical axes are for the solid and dashed curve, respectively.

In the decreasing width taper the group velocity of the solitons is continuously decreasing along the propagation length as shown in Figure 8(a), which causes the dispersive wave around 1.3 μm to eventually overtake the solitons. This raises the possibility of cross phase modulation (XPM) based spectral broadening of the dispersive wave [51]. In Figure 8(b) we see through numerical studies that the dispersive wave, which is initially slower than the rest of the pulse, eventually passes most of the signal around the length of 4 mm. Even though XPM is twice as effective as self-phase modulation, only a weak phase modulation of the dispersive wave ($\varphi_{max} \approx 1$) occurs due to the short interaction length (about 100 μm). Thus, only relatively weak spectral broadening takes place through XPM in this waveguide. Nevertheless, this raises the possibility of optimizing the design of the taper in future work, to increase the interaction length of the dispersive wave with the solitons thus increasing the XPM of the dispersive wave to broaden the short side spectrum even in the decreasing taper. We note that the signal always leading the pump pulse (in Figure 8b) is the long side dispersive wave and is undetectable by our OSA.

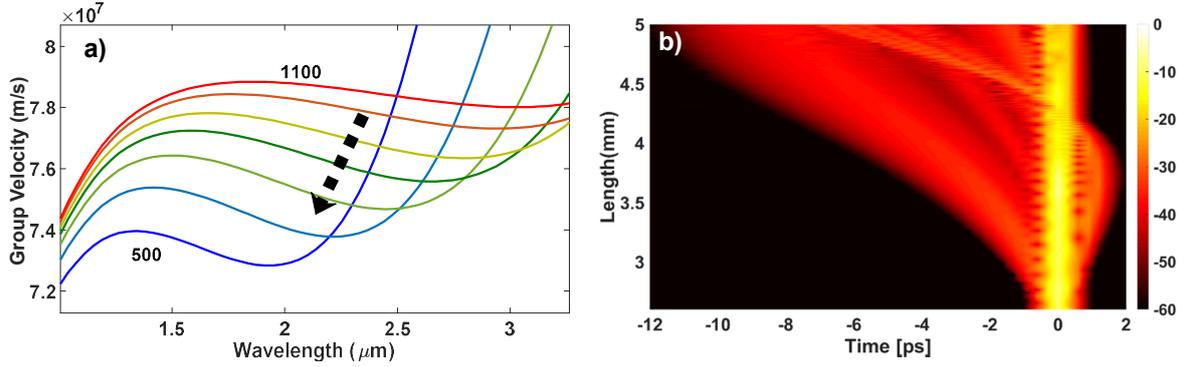

**Figure 8.** (a) The change in the group velocity along the length of the decreasing width taper indicated by the dashed arrow. The waveguide width is labeled above the curves. (b) The temporal evolution of the pulse in the decreasing taper (the pump pulse is centered at zero).

*3.2 Coherence in tapered continuum generation*
To determine the phase correlation between the modes of the SC signal, the coherence of the supercontinua generated in the tapers is simulated. In the simulation a pulse width of 250 fs and a peak power of 100 W is used. One photon per mode noise (shot noise) with 1.5% intensity noise was included in the pump, and the coherence was calculated for an ensemble of 100 supercontinua [52]. The coherence spectra are shown in Figure 9(a) for the increasing and decreasing width tapers and compared with that of a fixed width (920 nm) waveguide. The coherence of the supercontinua from the tapers is significantly higher than that from the fixed width waveguide. Along with the pulse compression [45], the coherence is improved due to the suppression of the modulation instability (MI) in the pulse in the tapers. The MI causes the generation of strong noise-based four-wave-mixing signals in the fixed width waveguide before the soliton fission takes place. The MI length of the fixed width waveguide is ~ 1.1 mm whereas, the soliton fission length is 1.5 mm. This in turn deteriorates the spectrum generated during the soliton fission process. The temporal evolution of the pulse near the soliton fission point for the increasing/decreasing taper and the fixed width waveguide are shown in Figure 9(b), (c) and (d). The temporal features of the pulses, with and without added noise in the simulations, are quite similar for the tapered waveguides, whereas for the fixed waveguide significant temporal modulation is present in the pulse due to noise. It is clear that the pulses in the fixed width waveguide are significantly deteriorated by the MI, whereas the pulses in the tapers are not affected by the noise. This happens because the dispersion, especially the GVD, is continuously changing in the taper thus the maximum MI gain frequency, given by [46] $\Delta\omega_{max} \propto P/(A_o\beta_2)$, where $A_o$ is the effective area, is continuously shifting. Therefore, an MI signal at a fixed frequency cannot build up enough strength to deteriorate the soliton fission process, unlike in the fixed width waveguide where the GVD is fixed.

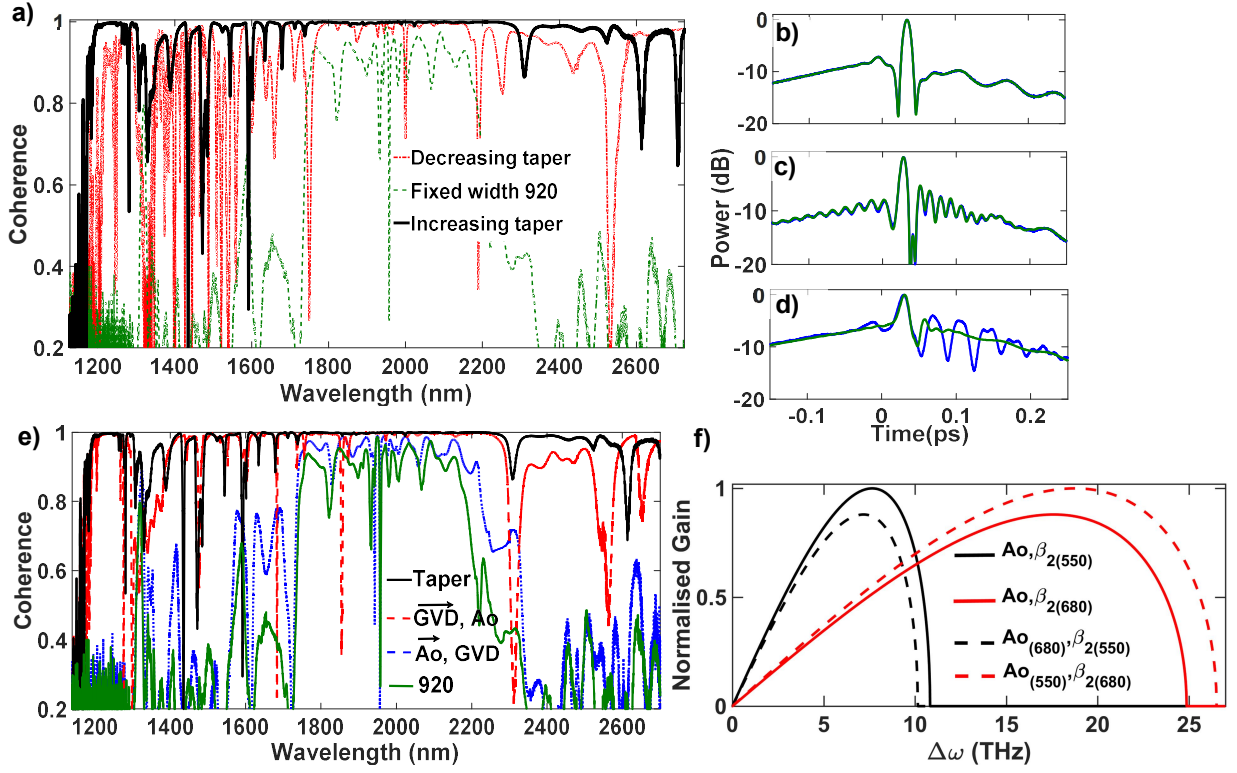

**Figure 9. (a)** The calculated coherence of the decreasing (red), increasing taper (black) and the fixed width 920 nm wide waveguide (green dash). **(b)**, **(c)**, and **(d)** The pulse at the soliton fission point with (blue) and without (green) adding noise in the simulation for the decreasing, increasing and fixed width waveguide. **(e)** The coherence of the increasing taper (black), waveguide with the varying GVD and fixed $A_o$ (red dash); waveguide with the varying $A_o$ and fixed GVD (blue dash), and the fixed width (920 nm wide) waveguide (green). **(f)** The normalized MI gain curves for the taper where its width is 550 nm (black) and 680 nm (red) are shown. The normalized MI gain curves for the GVD of 550 nm width and the $A_o$ of 680 nm width (black dashed), and vice versa (red dashed), are also shown.

The dependence of the normalized MI gain bandwidth on the effective area ($A_o$) and the GVD for 100 mW of peak power is shown in Figure 9(f). The gain bandwidth is calculated using the relation [46], $g = |\beta_2 \Delta\omega|(2\Delta\omega^2_{max} - \Delta\omega^2)^{1/2}$, where $\Delta\omega = \omega - \omega_{pump}$. We chose the variation in the effective area and the GVD for the increasing taper from the point where the taper width is 550 nm (close to input where dispersion is anomalous) to the fission point of 680 nm. We observe that the gain peak has shifted significantly between those two points of the taper. It is observed that the shift is more sensitive to variation in GVD than the variation in effective area (a similar trend is obtained with the full NLSE simulation including up to $8^{th}$ order dispersion terms and high peak power). This suggests that the effect of dispersion variation in avoiding the MI gain in the pulse is stronger than the effect of the effective area variation. To verify this, we calculated the coherence of the SC with fixed $A_o$ (GVD) and varying GVD ($A_o$), see Figure 9(e). It is observed that the SC coherence, by varying $A_o$ (with fixed GVD), is only slightly improved compared to the coherence of the fixed width waveguide, whereas, the variation of the GVD (with fixed $A_o$) improves the coherence significantly. Furthermore, we note that the coherence of the SC from the increasing width taper is slightly better than that of the decreasing width taper; that can be attributed to the shorter soliton fission length of the increasing width taper.

## 4. Birefringent waveguide

In this section the strong birefringence of the silicon waveguide is utilized to increase the supercontinuum bandwidth [38]. Even though the dispersion is not changing longitudinally one can utilize the two dispersions of an asymmetric waveguide due to the TE and TM modes. To demonstrate this, a fully etched strip waveguide was used with width of 1080 nm and height of 380 nm. The dispersion curves are shown in Figure 10(b) (inset) for the TE and TM modes.

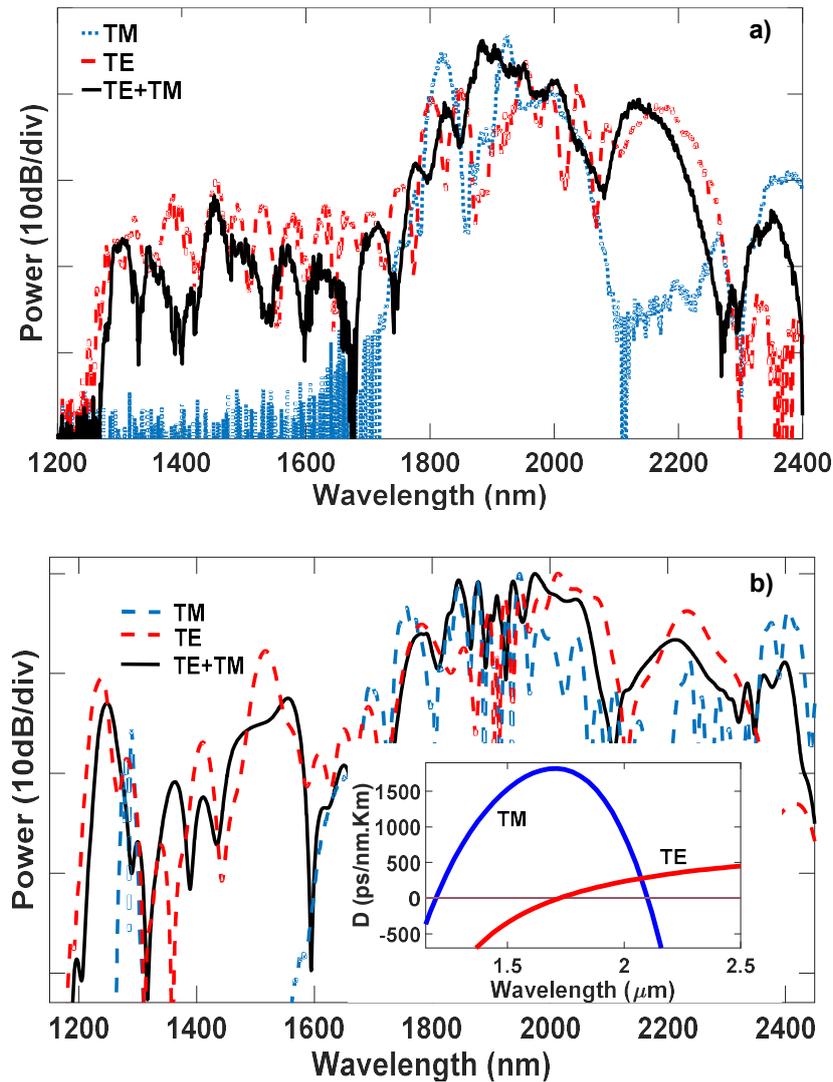

**Figure 10. (a)** Experimental and **(b)**, simulated supercontinua with the TM, TE and TE+TM (@ 45° to the plane of waveguide) mode pumping. The dispersion curves for the TE (red) and TM mode (blue), inset.

The experimental results of the supercontinuum generation are shown in Figure 10(a) for TM, TE and TM+TE (pump launched at 45° to the principle polarization axes). The pump pulse width is 250 fs, and the coupled peak power for TM and TE is 100 W, whereas for the TM+TE, the TE power is 70 W and the TM is 30 W, due to the un-optimized input inverse taper for the TM mode. The walk off length of the TE and TM modes is short (~150 μm), thus no intermodal coupling is expected. We observe in Figure 10(a) that the SC is broadened by >100 nm at the long wavelength side due to the linear superposition of the supercontinuum signals from the TE and TM mode pumping. The SC bandwidth can be increased further with this technique however we show the results up to 2.4 μm due to the limited bandwidth of our spectrum analyzer. Such a device offers a smaller footprint as it does not require separate waveguides for obtaining different dispersions as in ref. [42], as well as the capability for a broader SC based on dispersive wave generation [25].

**5. Discussions and Conclusion**
In this work, experimental and theoretical investigation of waveguides with varying-dispersion and birefringence for selective wavelength enhancement and improved coherence of the supercontinuum is reported. The results presented are independent of the material platform, so a wide variety of technologies should benefit. One must note that lower index contrast waveguides would require larger dimensional changes than the high-index contrast waveguides described here to achieve similar dispersion variation, thus compromising the footprint of the device. Although thickness variation of a waveguide can be an alternative for achieving dispersion variation, the fabrication can be challenging. In the decreasing taper, cross phase modulation can be utilized to spectrally broaden the dispersive waves where the dispersive waves temporally overlap not just with one fundamental soliton, as in the optical fiber, but with the entire signal in the anomalous regime. In the birefringent waveguide the input/output inverse tapers need to be optimized for coupling both modes efficiently. The TE and TM supercontinua could be rotated to be polarized as either TE or TM using the integrated polarization controllers [53].

In conclusion, with the work demonstrated here a robust all integrated system can be envisioned with an integrated mode-locked laser and germanium detector [54, 55] for signal synthesis that will have widespread applications in fields ranging from medicine to telecom.

**Acknowledgment**
This work was supported by Defense Advanced Research Projects Agency (DARPA) under the Direct on-chip digital optical synthesizer (DODOS) project—contract number HR0011-15-C-0056. NL acknowledges the NSS fellowship from Agency of Science, Technology, and Research (A*STAR), Singapore. Special thanks to Jelena Notaros for organizing the fabrication runs.

**Conflict of Interest**
The authors have declared no conflict of interest.

**Keywords**
Integrated photonics, Silicon photonics, Nonlinear photonics, Supercontinuum